\documentclass[]{spie}  
\usepackage[]{graphicx}
\usepackage[]{caption}
\usepackage[]{subcaption}
\usepackage[]{makecell}

\def\pasp{Publications of the Astronomical Society of the Pacific}

\def\aap{Astronomy \& Astrophysics}
\def\apj{Astrophysical Journal}
\def\procspie{Proc.~SPIE}
\def\aj{Astronomical Journal}
\def\araa{Annual Review of Astronomy \& Astrophysics}
\def\apjl{Astrophysical Journal Letters}
\def\mnras{Monthly Notices of the Royal Astronomical Society}
\def\apjs{Astrophysical Journal Supplement Series}

\title{Upgrading the Gemini Planet Imager: GPI 2.0}

\author{Jeffrey K. Chilcote\supit{a,b}, Vanessa P. Bailey\supit{c}, Rob De Rosa\supit{d}, Bruce Macintosh\supit{a}, Eric Nielsen\supit{a}, Andrew Norton\supit{a}, Maxwell A. Millar-Blanchaer\supit{c}, James Graham\supit{d}, Christian Marois\supit{e}, Laurent Pueyo\supit{f}, Julien Rameau\supit{g}, Dmitry Savransky\supit{h}, and Jean-Pierre Veran\supit{e}
\skiplinehalf
\supit{a}Kavli Institute for Particle Astrophysics and Cosmology, Stanford University, Stanford, CA 94305, USA; \\
\supit{b}Department of Physics, University of Notre Dame, 225 Nieuwland Science Hall, Notre Dame, IN, 46556, USA; \\
\supit{c}Jet Propulsion Laboratory, California Institute of Technology, 4800 Oak Grove Dr., Pasadena CA 91109, USA; \\
\supit{d}Department of Astronomy, UC Berkeley, Berkeley CA, 94720, USA; \\
\supit{e}National Research Council of Canada Herzberg, 5071 West Saanich Road, Victoria, BC V9E 2E7, Canada; \\
\supit{f}Space Telescope Science Institute, 3700 San Martin Drive, Baltimore MD 21218 USA; \\
\supit{g}Institut de Recherche sur les Exoplan{\`e}tes, D{\'e}partement de Physique, Universit{\'e} de Montr{\'e}al, Montr{\'e}al QC, H3C 3J7, Canada; \\
\supit{h}Sibley School of Mechanical and Aerospace Engineering, Cornell University, Ithaca, NY 14853, USA
}

\authorinfo{Further author information: (Send correspondence to Jeffrey K. Chilcote)\\Jeffrey K. Chilcote: E-mail: chilcote@stanford.edu}

\begin{document} 
  \maketitle 

\begin{abstract}
The Gemini Planet Imager (GPI) is the dedicated high-contrast imaging facility, located on Gemini South, designed for the direct detection and characterization of young Jupiter mass exoplanets. In 2019, Gemini is considering moving GPI from Gemini South to Gemini North. Analysis of GPI's as-built performance has highlighted several key areas of improvement to its detection capabilities while leveraging its current capabilities as a facility class instrument. We present the proposed upgrades which include a pyramid wavefront sensor, broadband low spectral resolution prisms and new apodized-pupil Lyot coronagraph designs all of which will enhance the current science capabilities while enabling new science programs.
\end{abstract}



\section{INTRODUCTION}
\label{sec:intro} 

Since the discovery of 51 Pegasi in 1995 \cite{MayorQueloz1995}, the search for and discovery of extrasolar planets has dramatically changed our understanding of planetary formation and the place where our solar system stands with respect to other solar systems in the galaxy. While thousands of exoplanets have now been discovered only a few of the most massive or widest separation have been observed spectroscopically. Direct imaging allows for the discovery of planets on solar system-scale orbits, provides new insight into the formation and characteristics of extrasolar systems, and enables direct spectroscopic observations of their atmospheres.

The Gemini Planet Imager (GPI) is a facility class instrument designed to address the fundamental goal of directly detecting and observing exoplanets. GPI was designed and built to directly image and spectroscopically characterize young, Jupiter-sized, self-luminous extrasolar planets and search for circumstellar debris disks that are sculpted by planetary systems. While designed for either Gemini North or Gemini South, GPI was installed at Gemini South in the fall of 2013\cite{Macintosh2014}. GPI finished commissioning and began science operations in the fall of 2014.

GPI consists of an adaptive optics (AO) system, apodized-pupil Lyot coronograph (APLC), a precision infrared wavefront sensor (CAL), and a near-IR integral field spectrograph (IFS). The AO system consists of a 4096-acuator micro-electro-mechanical (MEMS) deformable mirror, a CILAS 11 actuator diameter piezoelectric DM in a woofer-tweeter configuration, and a Shack-Hartmann WFS design with a Lincoln Labs CCID-66 sensor\cite{Macintosh2014,Poyneer2014}. GPI uses an apodized Lyot conograph (APLC) to suppress coherent light from the central star\cite{Soummer2011}. GPI's science instrument is an IFS with 192x192 spatial pixels dispersed through a prism to provide a resolving power of R$=\sim$30-100 depending upon the band. The GPI IFS has 5 individual filters in Y, J, H, and 2 in K-band (split into overlapping segments). The IFS further incorporates a Wollaston prism to allow for polarization measurements but only in broad band\cite{Larkin2014,Chilcote2012}.

In this work, we will present a brief overview of the current performance and major science results from GPI since 2013 (Section \ref{sec:CurrentGPI}). In section \ref{sec:nextgenscience} we will present a series of proposed new science goals for the next generation of GPI. Finally, we will present a series of proposed upgrades to GPI being considered -- called GPI 2.0 -- to enable it to best meet the presented science goals (Section \ref{sec:gpi2upgrades}).

\section{Summary of Major Science Results from GPI}
\label{sec:CurrentGPI}

GPI has been in routine operations at Gemini South since the 2014B semester. In addition to the large-scale exoplanet survey (GPIES), there have been two Large and Long Programs (LLP) which span multiple semesters, and 76 accepted regular proposals from 43 unique principal investigators. Twelve regular proposals were through Gemini's new Fast Turnaround program, which is designed for short high-impact programs. Based on their titles, about half of the accepted programs are related to exoplanet detection and/or characterization, a third are exploiting GPI's polarimetric capability to characterize circumstellar material around both young and more evolved stars. The remaining programs are focused on topics ranging from astrometric monitoring of young binaries to characterizing degenerate companions. Excluding time allocated for the GPIES campaign and science verification programs, 708 hours have been allocated to GPI observations between 2015A and 2018A, with ~310 hours having been executed. Additionally, 715 hours of GPIES campaign observations have also been taken between 2014B and 2018A which has operated exclusively in priority visitor mode since the start of the campaign.

\begin{figure}
\centering
\begin{subfigure}{.45\linewidth}
  \centering
  \includegraphics[width=\linewidth]{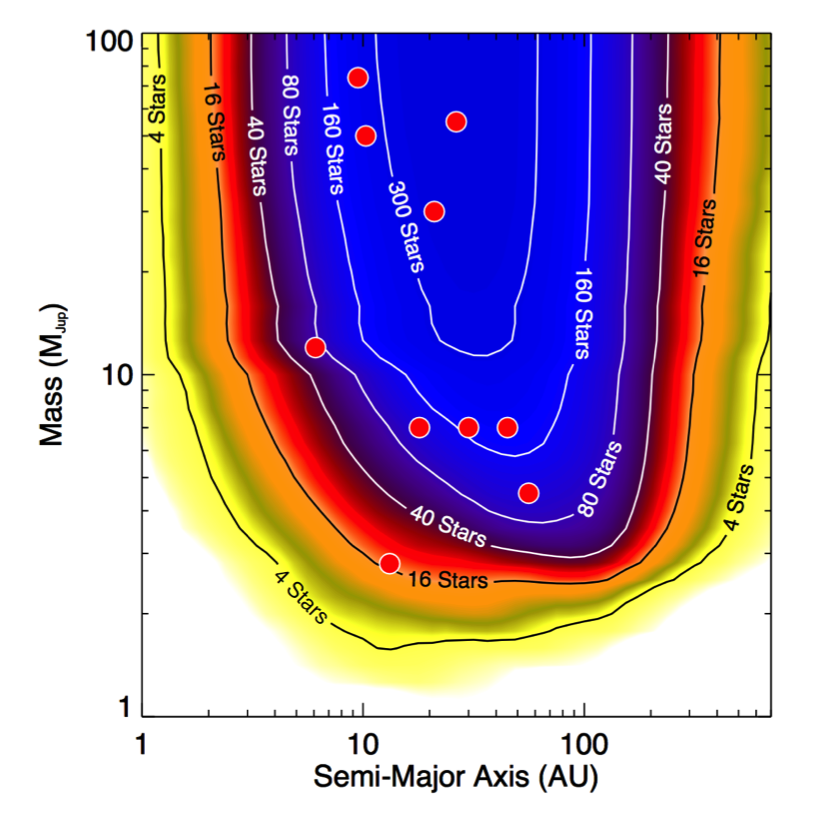}
\end{subfigure}%
\begin{subfigure}{0.45\linewidth}
  \centering
  \includegraphics[width=\linewidth]{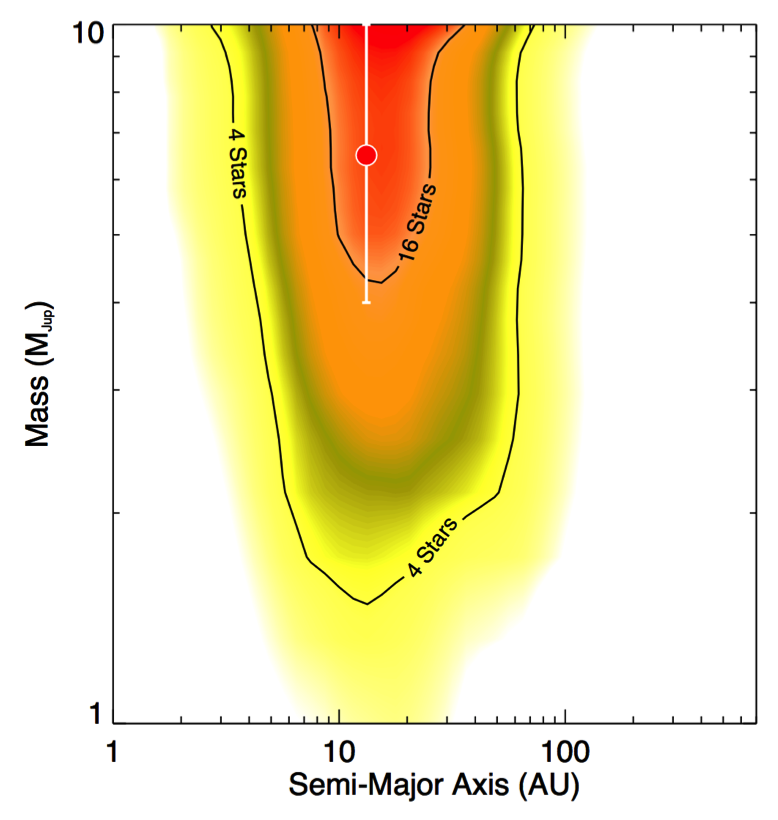}
\end{subfigure}
\captionsetup{width=.9\linewidth}
\caption{Sensitivity to hot \textbf{(Left)} and cold \textbf{(Right)} start planets from the GPIES campaign as a function of mass and semi major axis. Observed planets and brown dwarfs marked in red.} 
\label{fig:tongue}
\end{figure}

Since 2014, there have been 46 publications in peer-reviewed journals based wholly or in-part on GPI data, with a roughly even split between publications using data obtained during the GPIES campaign. One of the most impactful science results from GPI to date was the discovery of the exoplanet 51 Eridani b\cite{Macintosh2015}, the closest analog to Jupiter in terms of mass and semi-major axis amongst all directly imaged planets. GPI observations have been used to constrain the orbital parameters of the planet\cite{DeRosa2015} and the atmospheric composition and fundamental properties of the planet\cite{Rajan2017}. In addition to the discovery of a new brown dwarf companion to HR 2562\cite{Konopacky2016}, GPI has been used to spectroscopically characterize previously known substellar companions\cite{Ingraham2014,Galicher2014, DeRosa2016, Chilcote2017, Johnson-Groh2017, Greenbaum2018, Crepp2018}, providing new insights into the diversity of substellar companion atmospheres at a range of masses and ages. The architectures of planetary systems have also been investigated, with GPI observations providing indirect evidence of the presence of additional companions within the HD 95086 system\cite{Rameau2016}, and demonstrating that the Hill sphere of $\beta$ Pictoris b would transit its host star in 2017-18\cite{wang2016}. Statistical results from the first half of the GPIES survey are to be presented in an upcoming paper (Nielsen et al. 2018, in prep), placing some of the best constraints to-date on the frequency of wide-orbit giant planets. GPI's spectral and polarimetric capabilities have also been used to search for and characterize transition disks\cite{Reggiani2014, Currie2015, Rapson2015, Long2017, Follette2017} and debris disks\cite{Kalas2015, Draper2016, Millar-Blanchaer2016} around nearby young stars. These observations have been use to determine the geometry and composition of circumstellar disks\cite{Perrin2015} and postulate the presence of unseen planetary-mass companions based on asymmetries or structure resolved within the disk\cite{Esposito2016, DongFung2017}.

\begin{figure}
\centering
  \includegraphics[width=0.8\linewidth]{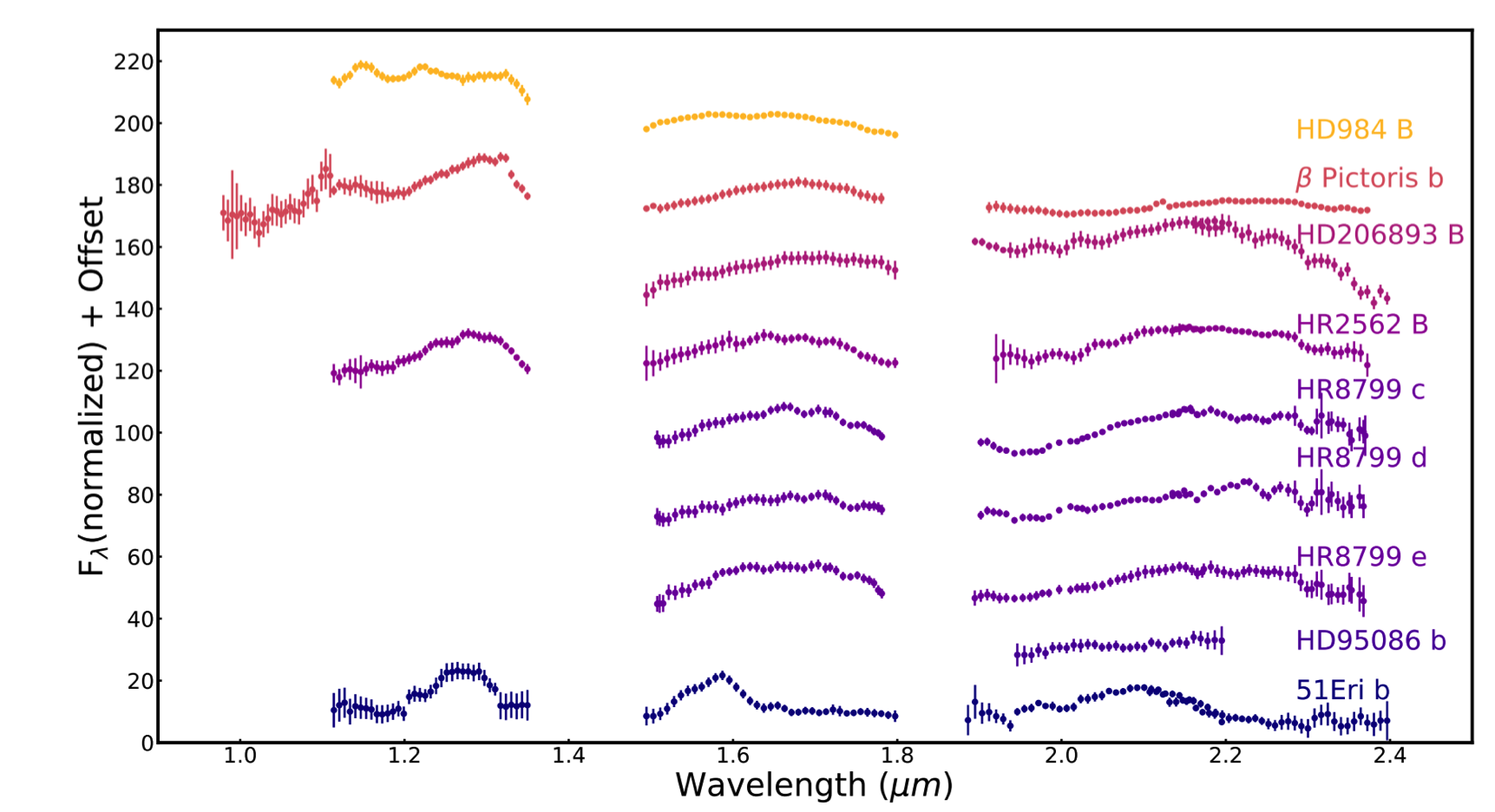}
  \caption{A library of published spectra as observed with GPI.}
  \label{fig:GPI_Spectrum_library}
\end{figure}

\begin{figure}
\centering
\begin{subfigure}{.47\linewidth}
  \centering
  \includegraphics[width=\linewidth]{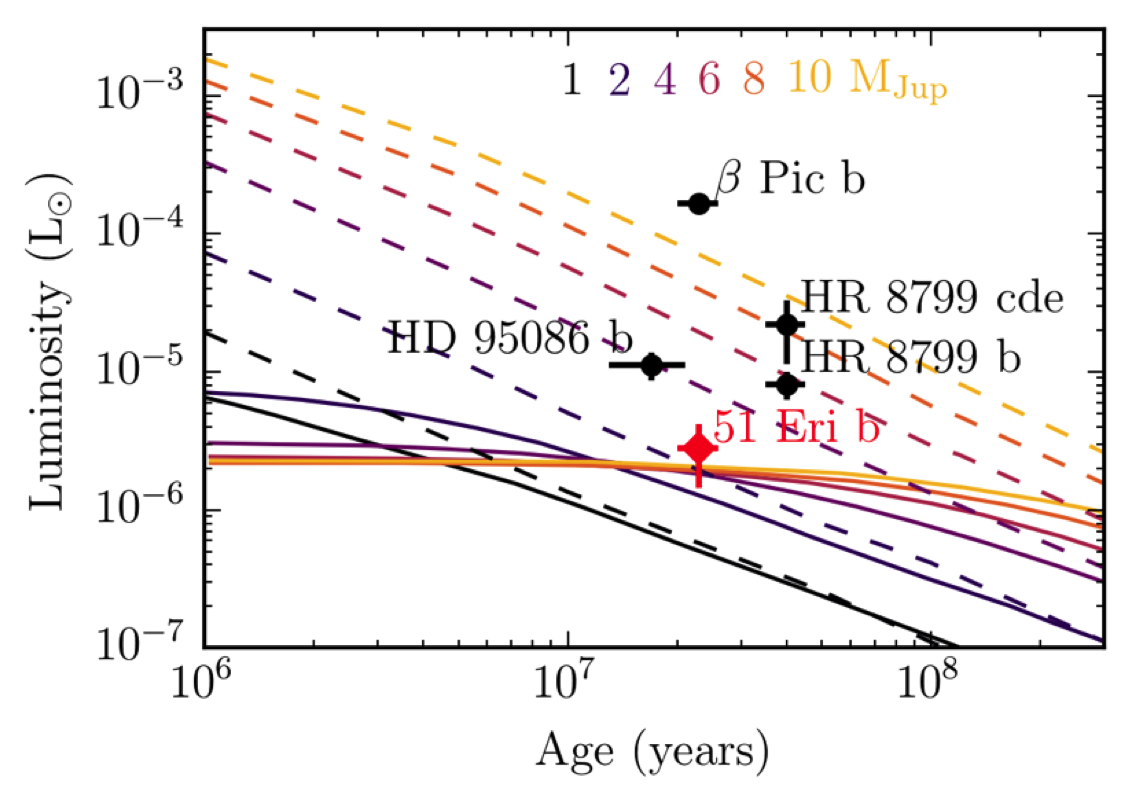}
\end{subfigure}%
\begin{subfigure}{0.3\linewidth}
  \centering
  \includegraphics[width=\linewidth]{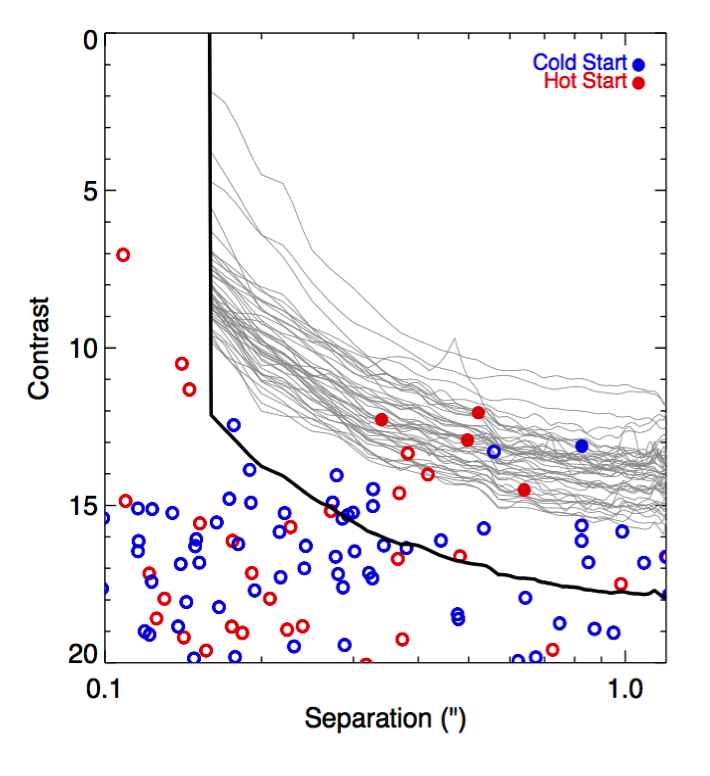}
\end{subfigure}
\captionsetup{width=.9\linewidth}
\caption{\textbf{Left:} Hot-start (dashed) vs cold-start evolutionary models for 1-10Mj with age vs luminosity\cite{Marley2007}. Several known planets indicated. \textbf{Right:} In gray are typical GPI contrast curves with the present instrument. In black is a prediction of improving GPI's contrast by 3.2 magnitudes. The planet population is modeled using a distribution of planets consistent with radial velocity and direct imaging surveys.  Filled circles represent planets that could be detected with the current observations, open circles would fall below the contrast curve for their particular host star.} 
\label{fig:planet_population}
\end{figure}

\section{Next Generation Science Goals}
\label{sec:nextgenscience}
Currently under review is the possibility of moving GPI from Gemini South to Gemini North sometime in 2019\cite{Macintosh2018,Rantakyro2018}. GPI has been in nearly continuous operation since 2013 without a significant maintenance overhaul to its internal components. Any move from Chile to Hawaii would require significant disassembly of the instrument and an inspection of its internal components to ensure the safety and reliability of the instrument as part of a move. This relocation and maintenance cycle would be an ideal time to improve the instrument with hardware upgrades to address the next generation of science requirements. There are several areas of science that might benefit from improvements to GPI's capabilities, which are within the technical capabilities of an upgrade.

\subsection{Large Scale Survey \& Cold-start planets}

An active area of study for young, giant exoplanets is how the current luminosity of these young planets traces their formation mechanisms, and whether these planets can be described by a ``hot start'' luminosity model (likely corresponding to a disk instability formation mechanism), or the alternate ``cold start'' models, which is expected from cold start formation.  The current GPIES survey from Gemini South is quite sensitive to hot-start giant planets between 10-100 AU, but the survey sensitivity is comparatively low to cold-start planets (Figure~\ref{fig:tongue}).  Indeed, based on current cold-start evolutionary models\cite{Fortney2008} coupled with AMES-COND atmosphere models\cite{Allard2001,baraffe2003}, nine of the ten planets and brown dwarfs detected by the campaign can only be hot start objects, while only 51 Eridani b could potentially be a cold-start planet.  If hot start, 51 Eri b would be 2.7 M$_{Jup}$, if cold start it can be between 4 and 10 M$_{Jup}$.  This is a direct consequence of the significantly lower luminosity ($\sim 2 \times 10^{-6}$L$_{\odot}$) predicted for young ($\leq$ 100 Myr) cold start planets compared to their hot start counterparts (Figure~\ref{fig:planet_population}, left panel).

A modest improvement in the contrast reached by GPI, however, can dramatically increase its sensitivity to cold start planets.  As seen in Figure~\ref{fig:planet_population}, while cold start models predict low-luminosity planets, the overlap of models at $\sim 2 \times 10^{-6}$L$_{\odot}$ means a small increase in achieved contrast around a young star will rapidly switch from no sensitivity to cold start planets to being sensitive to cold start planets between 2 and 10 M$_{Jup}$.  This is a stark difference from hot start planets, where luminosity increases gradually and monotonically with mass.  The right panels of Figure~\ref{fig:planet_population} shows a possible population of giant planets around GPIES stars consistent with our observations, as a function of contrast with the host star in magnitudes and separation from the star.  While improving contrast by 3.2 magnitudes would lead to a moderate increase in yield in hot start planets, it would increase the yield of cold-start planets by a factor of 10.

\begin{figure}
\centering
\begin{subfigure}{.45\linewidth}
  \centering
  \includegraphics[width=\linewidth]{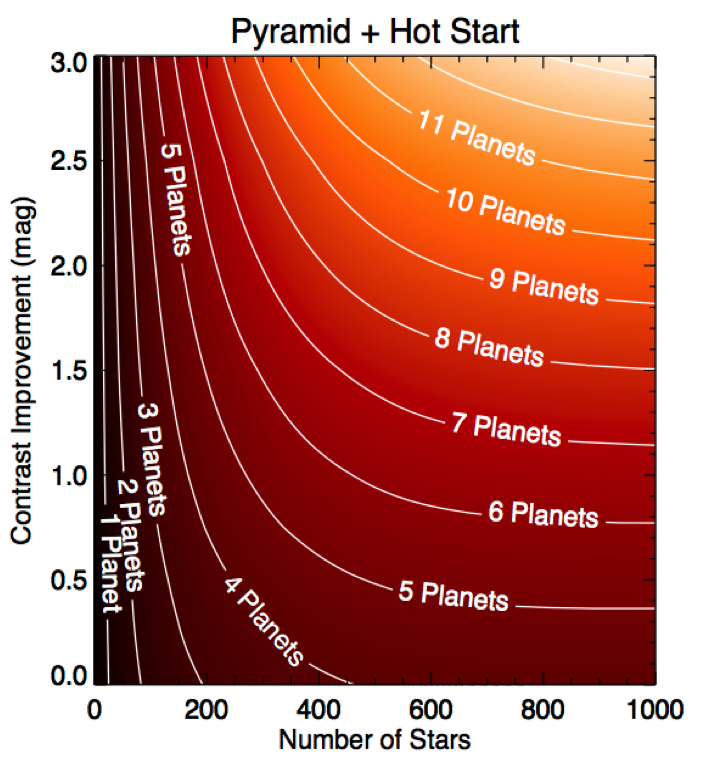}
\end{subfigure}%
\begin{subfigure}{0.45\linewidth}
  \centering
  \includegraphics[width=\linewidth]{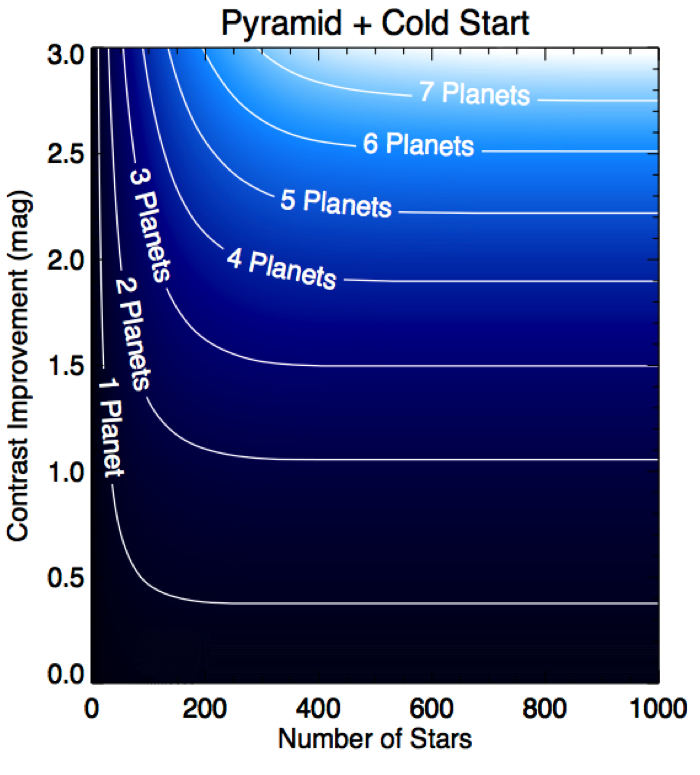}
\end{subfigure}
\captionsetup{width=.8\linewidth}
\caption{The number of new hot \textbf{(Left)} and cold \textbf{(Right)} start planets discovered as a function of contrast improvement and survey size for a survey from Gemini North.} 
\label{fig:hot_cold_start_survey}
\end{figure}

If we couple the improvement in instrument contrast with a moderate-scale survey ($\sim$200 stars) at Gemini North, we can dramatically increase the GPI planet yield for a tailored survey making efficient use of observing time.  We have compiled a list of 1643 young, nearby stars accessible from Gemini North that have not been observed by GPIES on Gemini South, and utilize a model of planet populations consistent with the current results from the GPIES campaign.   Figure~\ref{fig:hot_cold_start_survey} shows predicted planet yield as a function of number of stars in the survey (stars are ordered so the best stars are observed first), as a function of improvements in the contrast (0 improvement represents GPI as-is).  So, conducting a 200-star northern survey with GPI would be expected to yield 3 hot start planets, while a 1.5 magnitude contrast improvement would increase the yield to 5 planets.  The gain is more pronounced for cold start planets, where the planet yield would increase from 0 expected cold start planets to 3 for the same survey, if cold start and hot start planets follow the same underlying distributions.  We note that there are no known planets around the stars in our target list, so these planets would represent new discoveries.

For modest improvements in the contrast, and a moderate investment in observing time compared to other direct imaging surveys, GPIES 2.0 is poised to make a significant increase in the number of cold-start and hot-start planets, allowing us to probe the formation mechanism of these wide-separation giant planets for the first time.

\subsection{Very young stars \& transitional disks}
Beyond planets around young stars in general, there is interest in looking at planets in the epoch of formation. These are solar systems and planets actively forming and recently formed around stars less then 1 million years old\cite{Luhman2010}. For an observatory located in the northern hemisphere, the majority of these solar systems would be observed in Taurus. These planets would be extremely bright\cite{Marley2007,Burrows1997}, allowing GPI to see to even the lowest mass planetary companions. If GPI were able to achieve a brightness limit of V$\sim12$th magnitude there would be approximately 60 targets available for such a survey. 

These targets would be ideal for observing transitional disks due to the age of the systems. Transition disks are circumstellar disks that still have a substantial gas component (unlike debris disks) but also a central clearing (unlike protoplanetary disks). Transition disks are often considered sites of ongoing planet formation, representing a critical time in the coevolution of disks and planets. This central cavity is generally a few to a few tens of AU in size. The Northern hemisphere hosts many more known transitional disks than the Southern hemisphere. 

For both of these projects, looking at similar sets of stars, an extremely close inner working angle would be required for planets and disks due to the distance of Taurus ($\sim140$pc)\cite{Luhman2010} as well as the ability to achieve a fainter operating magnitude (e.g. V=$\sim13$th).

\subsection{Spectropolarimetry}
Comparisons between directly imaged planet (and brown dwarf) spectra and state of the art atmospheric models have revealed systematic offsets that are often attributed to our lack of detailed understanding of the cloud properties\cite{MarleyRobinson2015,Rajan2017}. In this respect, polarimetry can be a powerful tool as it is extremely sensitive to the properties of atmospheric scatterers, such as cloud height, patchiness and composition\cite{Marley2011,Stolker2017}. However, with broadband polarimetric measurements alone, many of these properties remain degenerate. Higher resolution spectropolarimetry is able to break some of these degeneracies, by measuring the degree of linear polarization in spectral absorption bands, providing critical information on the optical depth of absorbers below the cloud decks\cite{deKok2011}. Thus spectropolarimetry is able to provide important contraints on cloud properties in directly imaged planets that are unavailable through other observational techniques. Other science cases include spectropolarimetric measurement of dust scattering properties in debris disks, which can be more reliable than comparing non-contemporaneous broadband measurements, as well as spectropolarimetric measurements of clouds in solar system objects, such as Neptune. Making these measurements with GPI would require only a modest upgrade as the spectral resolution needed to make these measurements matches well the exisiting spectroscopic capabilities of GPI and much of the existing polarimetric system could be re-used. Spectropolarimetric observations of directly imaged planets will be highly complemented by the on-going spectropolarimetric survey of  free-floating brown dwarfs being carried out with the WIRC+Pol instrument at Palomar observatory.

\begin{table}
\centering
\def\arraystretch{2.5}
\resizebox{0.95\textwidth}{!}{
\begin{tabular}{lccc}
\textbf{Science Cases} & {\textbf{WFS I mag limit}} & {\textbf{Inner working angle}} & \textbf{Additional Improvement} \\
\hline
\makecell[l]{Large Scale Survey \& \\Cold-start planets} & 10    & $0.15''$  & 2+ mag contrast\\
\makecell[l]{Very young stars \& \\ transitional disks} & 13    & $0.1''$   & 0 \\
Spectropolarimetry & 7 & $0.15''$ & 1\% polarimetry \\
Low Mass Stars & 13-14 & $0.1''$ & 0 \\
\makecell[l]{Asteroids \& \\ Solar System Objects} & 14 & - & 0 \\
Debris Disks & 9     & $0.2''$   & 0 \\
\makecell[l]{Planet Variability \& \\abundance characterization} & 6 & $0.2''$ & 1\% photometry
\end{tabular}
}
\vspace{0.1in}
\caption{Proposed GPI science cases and system performance requirements to achieve science cases.}

\end{table}

\subsection{Low Mass Stars}
M-stars are the most numerous stars in our galaxy. They offer a unique environment to understand the formation of planetary systems, their formation mechanisms, and to find the lowest mass planets due to M-stars lower flux. Around M-stars gravitational instability is extremely efficient at forming planets providing a test for gravitational instability\cite{KennedyKenyon2008}. Unfortunately, most radial velocity surveys stop around $\sim3$AU with direct imaging surveys stopping around $\sim30$AU\cite{Bowler2016}. Additionally, these systems host interesting debris disks (e.g. AU Mic). These systems are typically fainter requiring an AO system to achieve high correction for faint stars. In order to probe these systems, GPI would need to achieve $\sim$ 14th mag for $\sim50$ targets. 

\subsection{Asteroids \& Solar System Objects}
Asteroids in our solar system are metallic, rocky and icy objects ranging in size from a few meters to a few hundred kilometers. While we have constraints for the surface composition, albedo and rotation rates for many main-belt asteroids the 3-D shape, crater distribution and density have only been measured for a limited number of bodies. Characterizing these physical properties would allow us to address entirely new questions regarding the earliest stages of planetesimal formation and their subsequent collisional and dynamical evolution. By operating GPI at a lower visible magnitude limit for its AO system while maintaining high contrast capabilities will allow expansion of the number of targets studied. For example, if the AO system were able to achieve a limiting magnitude of V=14 then $\sim1300$ objects would be available for study. GPI brings with it the ability to provide spectroscopy allowing for an observation of the mafic bands to characterize the surface composition. Further, GPI's AO capabilities are needed in order to study the properties of companion asteroids, not only single asteroids.

\subsection{Planet Variability}
Understanding the atmospheres of exoplanets is one of the key goals of direct imaging. Current models of the characterization of exoplanets such as HR8799c,d,e and $\beta$~Pic~b indicate the presence of both non-equilibrium chemistry and patchy clouds, which can confound inferences of their composition\cite{Skemer2014}. Patchy clouds are often indicated as important factors in understanding the history and composition of these planetary atmospheres. As planets rotate, variable cloud cover induces modulation in the flux of a planet. This variability has been used to measure the rotation rates and study the atmospheres of planetary-mass bodies and brown dwarfs. In order to study this variability for a significant number of targets, it is estimated that GPI would need to improve its capabilities to achieve 1\% photometric stability on the brightest targets.

\begin{figure}
\centering
\begin{subfigure}{.45\linewidth}
  \centering
  \includegraphics[width=\linewidth]{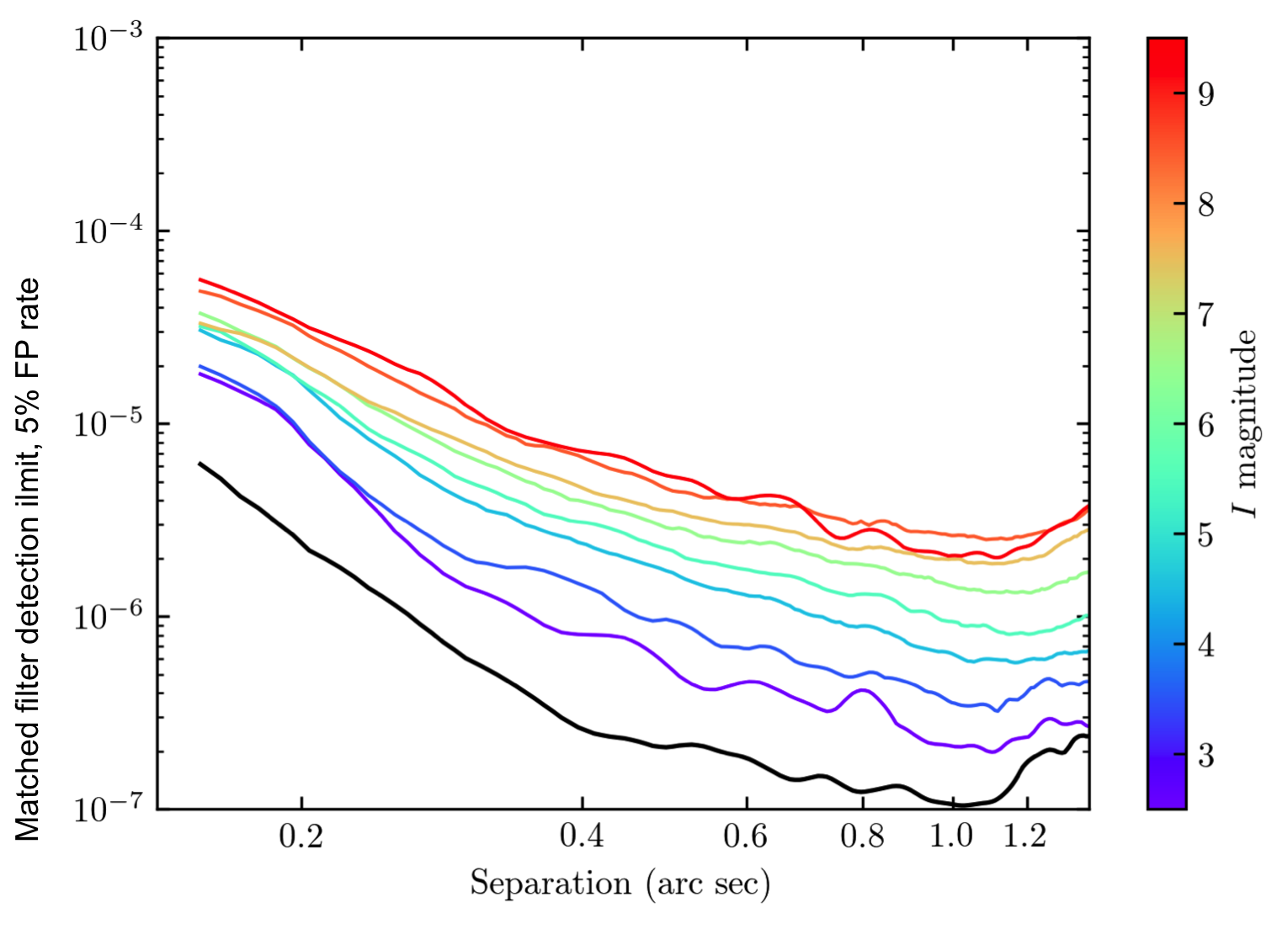}
\end{subfigure}%
\begin{subfigure}{0.45\linewidth}
  \centering
  \includegraphics[width=\linewidth]{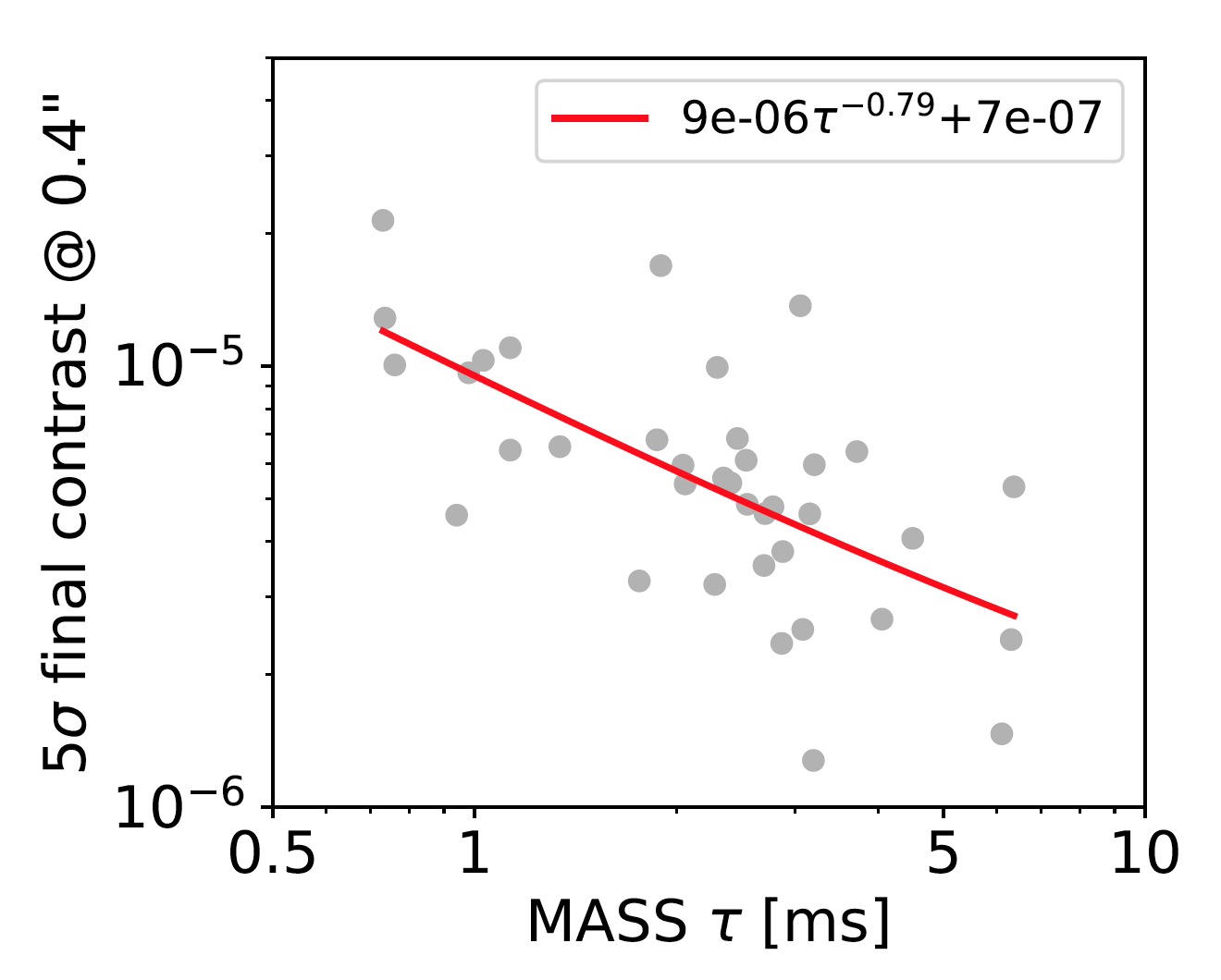}
\end{subfigure}
\captionsetup{width=.8\linewidth}
\caption{\textbf{Left:} GPI final contrasts, 0.05 false positives per epoch, as a function of guide star I magnitude as computed with Forward Model Matched Filter (FMMF)\cite{Ruffio2017}. The contrast does not seem to have reached a performance floor as the brightness of GPI improves, but as brightness improves the contrast continues to improve indicating that GPI has not yet reached a systematic floor. \textbf{Right:} GPI measured performance as a function of coherence time. A faster AO system will behave as though the seeing was slower along with slower Gemini North conditions\cite{Rantakyro2018}.} 
\label{fig:AO_Performance}
\end{figure}

\section{GPI 2.0 Instrument Upgrades}
\label{sec:gpi2upgrades}
To fulfill these science goals, a number of upgrades to GPI have been proposed. These will address both expanded science goals for GPI (Section \ref{sec:nextgenscience}) and improve GPI based upon lessons learned from GPI and other instruments. The following is a list of upgrades currently under evaluation for an upgrade to GPI in order to allow it to meet the next generation of science goals. 

\begin{itemize}
    \item Replace GPI's Shack-Hartman WFS (SHWFS) with a Pyramid WFS (PWFS). Testing by the Thirty Meter Telescope's (TMT's) NFIRAOS AO team has shown that using a PWFS instead of a SHWFS allows them to gain $\sim1.3$ magnitudes for mostly all reference star brightness cases\cite{Veran2015}. This will allow GPI to not only operate at a lower magnitude, but will allow it to maintain a better correction than it is currently capable of as stars become fainter.
    
    \item Replace GPI's Lincon Labs CCD with an EMCCD. GPI's Lincon Labs CCID-66 sensor has achieved $\sim4$ electrons rms noise at 1172 Hz frame rates, and in practice achieves  $\sim6.7$ electrons rms noise in standard operations. Further, the CCID-66 has a quantum efficiency of $\sim70\%$. Switching to a EMCCD such as an OCAM2K will enable GPI to maintain operational capacity another $\sim2$ magnitudes fainter and will allow for faster operations of the WFS on brighter targets.
    
    \item Replace the GPI AO Real Time Computer (RTC) with an updated computer enabling a decrease in the computation time. GPI's measured performance is shown as a function of coherence time in figure~\ref{fig:AO_Performance}. As the seeing slows down, GPI's performance is improving and no floor has yet been identified. A faster AO system will behave as though the seeing was slower even under the same conditions.

    \item As demonstrated by GPI's TT LQG, advanced control methods can make up for temporal delay\cite{Poyneer2014}. If GPI can implement an explicitly predictive scheme for high-order phase errors this will effectively slow down the atmosphere improving GPI's contrast capabilities. Multiple schemes are being investigated for this including but not limited to Fourier-based wind predictors.
    
    \item GPI was designed to only offload slow movements to the Gemini secondary. In operations GPI has difficulty with its Tip-Tilt range in higher wind conditions. As part of an upgrade, we are exploring increasing the range of the AO system's Tip-Tilt mirror and increasing the update rate to the Gemini secondary as possible solutions to tracking wind shake and allowing GPI to operate in higher wind conditions.
    
    \item For photometric calibration, GPI currently uses a fixed grid in the pupil plane as part of the APLC apodizer. Due to the coherence with the speckle field, accuracy of the ``satellite-spots'' generated via this method is limited. By moving to time-averaged incoherent speckles generated by the DM, we hope to improve the accuracy of the satellite-spots enabling improvements in the overall image photometry\cite{Jovanovic2015}.
    
    \item GPI's APLC uses designs from modeling and thought processes which are nearly 10 years old. Using the latest APLC designs it is believed a better contrast can be achieved close to the star. Further, because the latest designs upgraded contrast deeper close to the star and the focal plane mask is chosen to be more robust than previous generations the contrasts degrades less quickly with tip-tilt jitter RMS (Figure~\ref{fig:APLC_Performance}).
    
    \item To help deal with quasi-static speckles to reach $\sim3$ fainter under consideration is a fast focal plane wavefront sensor module\cite{Gerard2018}.

\begin{figure}
\centering
  \includegraphics[width=0.5\linewidth]{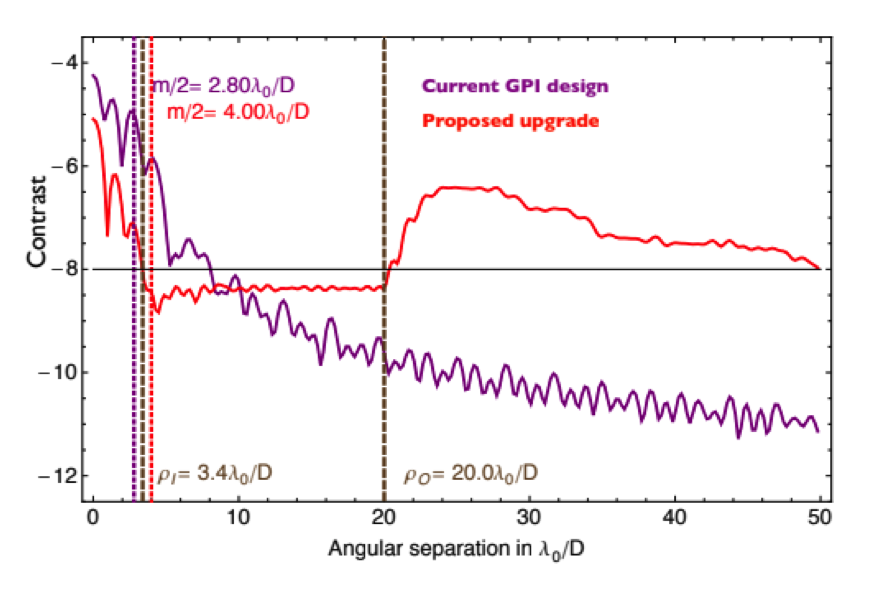}
  \captionsetup{width=.4\linewidth}
  \caption{Updated APLC designs assuming a perfect input wavefront\cite{NDiaye2015}. 2.8$\lambda$/D = $\sim40$mas in GPI H-band.}
  \label{fig:APLC_Performance}
\end{figure}

    \item GPI's IFS \& CAL use Goodrick Sensors Unlimited Near-IR Area cameras. Upgrading these to CRED-2 cameras will increase the speed and noise performance of the CAL for tracking the star behind the coronagraph and for improving pupil alignment accuracy in the IFS.
    
    \item GPI provides operation in 5 individual filter bands Y,J,H, and K (split into 2 filters)\cite{Larkin2014,Chilcote2012}. In order to achieve a higher SNR on dim stars and improve inner working angle GPI is investigating incorporating a low-resolution Y-K band mode similar to the mode provided in the CHARIS instrument on SCExAO\cite{Groff2016}. This will further allow for GPI to understand the photometric differences between bands and better calibrate photometry and satellite  spots.
    
    \item Currently GPI can only preform spectroscopy or polarimetry but not both. A number of design changes are being considered to enable spectroscopy and polarimetry simultaneously. Using the existing filter set along with a low resolution mode of R$=\sim20$ would enable low resolution spectro-polarimetry by shortening the current spectral length providing the needed space for simultaneous polarization measurements on the same H2RG currently in GPI. A second option would be to provide a 2nd H2RG in GPI to allow for simultaneous polarization measurements while maintaining all the currently available spectroscopy modes.
    
    \item GPI has only been able to achieve about 5\% photometric accuracy. While this is good, it is not enough to understand the cloud structure of planets to learn about their rotation rates. Pipeline improvements and the above DM sat spot improvements are being investigated to see if GPI can achieve 1\% photometric accuracy on the brightest stars. 

\end{itemize}

\section{Conclusion}

Many of the design decisions which went into the construction of GPI were made in 2004. The technology and field of ExAO systems has greatly advanced in the intervening years. Since GPI started standard science operations on sky in 2014, it has not had any major hardware changes to the instrument. The lack of hardware changes had lead to the stability and reliability that has helped GPI to achieve its major science goals and to provide a platform to cutting edge science. In order to stay competitive for the next decade, GPI should take into account the significant improvements made in the field and technology. In this work, we have laid out a number of proposed science cases which GPI should work to address. The team is currently studying a number of hardware upgrades aimed at addressing these science cases. The move from Gemini South to Gemini North is an ideal time to implement these hardware changes and upgrade GPI to enable cutting edge science from Gemini North for many years to come.

\acknowledgments     
 The research was supported by grants from NSF AST-1411868 and NASA NNX14AJ80G. Research benefitted from the Gemini Observatory, operated by AURA for NSF and the Gemini consortium also NNX15AD95G and NNX15AC89G . This work benefited from NASAs Nexus for Exoplanet System Science (NExSS) research coordination network sponsored by NASAs Science Mission Directorate. V.B. acknowledges government sponsorship; this research was carried out in part at the Jet Propulsion Laboratory, California Institute of Technology, under a contract with the National Aeronautics and Space Administration.


\bibliographystyle{spiebib}   

\end{document}